\documentclass[12pt]{article}
\usepackage[dvips]{graphicx}
\def\doublespace{\lineskip      .25 ex\baselineskip 3.0
ex\lineskiplimit 0 ex\parskip 1.0 ex plus.50 ex minus .25 ex}%

\begin{document}
\doublespace

\title{Braneworld Stars: Anisotropy \\ Minimally Projected Onto the
Brane}

\author{ J.
Ovalle\footnote{jovalle@usb.ve}
\footnote{Based on the contributed lecture given at the IX Asia-Pacific International Conference on Gravitation and Astrophysics (ICGA9), June 28-July 2, 2009, Wuhan, China.}
 \\
\vspace*{.25cm}\\
 Departamento de F\'{\i}sica, \\ Universidad Sim\'on Bol\'{\i}var, \\ Caracas,
 Venezuela.
 }
\date{}
\maketitle
\begin{abstract}

In the context of the Randall-Sundrum braneworld, an exhaustive and detailed description of the approach based in the
minimal anisotropic consequence onto the brane, which has been
successfully used to generate exact interior solutions to
Einstein's field equations for static and non-uniform braneworld
stars with local and non-local bulk terms, is carefully presented. It is shown that this approach allows the generation of a braneworld version for any known general relativistic solution.

\end{abstract}
\section{Introduction}
Is well known that the non-closure of the braneworld equations represents an open problem in the study of braneworld stars \cite{maartRev2004}-\cite{gergely2007}. A better understanding of the bulk geometry and proper boundary conditions is required to overcome this issue. Since the source of this problem is directly related with the projection ${\cal
E}_{\mu\nu}$ of the bulk Weyl tensor on the brane, the first logical step to overcome this issue would be to discard the cause of the problem, namely, to impose the constraint ${\cal
E}_{\mu\nu}=0$ in the brane. However it was shown in \cite{koyama05} that this condition is incompatible with the Bianchi identity in the brane, thus a different and less radical restriction must be implemented. In this respect, a useful path that has been successful used consist in to discard the anisotropic stress associated to ${\cal
E}_{\mu\nu}$, that is, ${\cal P}_{\mu\nu}=0$. However, in our opinion, this constraint, which is useful to overcome the non-closure problem \cite{shtanov07}, represents a restriction too strong in the brane. The reason is that some anisotropic effects onto the brane should be expected as a consequence of the "deformation" undergone by the 4D geometry due to five dimensional gravity effects, as was clearly explained in \cite{jovalle207}.

As was already shown in \cite{jovalle07} for braneworld stars, a useful constraint arises on the brane when the general relativity limit is demanded on any braneworld solution. There is a particular solution to this constraint, a constraint itself in the brane, the physical meaning of which represents a condition of minimal anisotropy projected onto the brane. In this paper, it is shown that this condition not only ensures a correct low energy limit, but also it represents a condition that is satisfied for {\it any} known general relativistic solution.

The principal goal of this paper is to show that demanding the minimal anisotropic effects onto the brane, which is explained in detail in the present work, it is possible to construct the braneworld version of every general relativistic solution, thus overcoming the non-closure problem of the braneworld equations by a natural braneworld generalization of any general relativistic solution.

\section{Non locality and the general relativity limit problem.}

 The effective Einstein's field equation in the brane can be written as a modification of the standard field equation through an energy-momentum tensor carrying bulk effects onto the brane:
\begin{equation}\label{tot}
T_{\mu\nu}\rightarrow T_{\mu\nu}^{\;\;T}
=T_{\mu\nu}+\frac{6}{\sigma}S_{\mu\nu}+\frac{1}{8\pi}{\cal
E}_{\mu\nu},
\end{equation}
here $\sigma$ is the brane tension, with  $S_{\mu\nu}$ and
$\cal{E}_{\mu\nu}$ the high-energy and  non-local corrections respectively.

Using the line element in Schwarzschild-like coordinates
\begin{equation}
\label{metric}ds^2=e^{\nu(r)} dt^2-e^{\lambda(r)} dr^2-r^2\left( d\theta
^2+\sin {}^2\theta d\phi ^2\right)
\end{equation}
in the case of a spherically symmetric and static distribution having Weyl stresses in the interior, the effective equations can be written as
\begin{equation}
\label{usual} e^{-\lambda}=1-\frac{8\pi}{r}\int_0^r r^2\left[\rho
+\frac{1}{\sigma}\left(\frac{\rho^2}{2}+\frac{6}{k^4}\cal{U}\right)\right]dr,
\end{equation}
\begin{equation}
\label{pp}\frac{8\pi}{k^4}\frac{{\cal
P}}{\sigma}=\frac{1}{6}\left(G_1^1-G_2^2\right),
\end{equation}
\begin{equation}
\label{uu}\frac{6}{k^4}\frac{{\cal
U}}{\sigma}=-\frac{3}{\sigma}\left(\frac{\rho^2}{2}+\rho
p\right)+\frac{1}{8\pi}\left(2G_2^2+G_1^1\right)-3p
\end{equation}
\begin{equation}
\label{con1}p_{1}=-\frac{\nu_1}{2}(\rho+p),
\end{equation}
with
\begin{equation}
\label{g11} G_1^1=-\frac 1{r^2}+e^{-\lambda }\left( \frac
1{r^2}+\frac{\nu _1}r\right),
\end{equation}
\begin{equation}
\label{g22} G_2^2=\frac 14e^{-\lambda }\left[ 2\nu _{11}+\nu
_1^2-\lambda _1\nu _1+2 \frac{\left( \nu _1-\lambda _1\right)
}r\right].
\end{equation}
where $f_1\equiv df/dr$ and $k^2=8{\pi}$. The general relativity
is regained when $\sigma^{-1}\rightarrow 0$ and (\ref{con1})
becomes a lineal combination of (\ref{usual})-(\ref{uu}).

As it is clearly shown through the Eqs. (\ref{usual})-(\ref{con1}), in the case of a non-uniform static distribution with local bulk
terms (high energy corrections) and non-local bulk terms (bulk
Weyl curvature contributions), we have an indefinite system of
equations in the brane, that at the end is represented by the set of three unknown functions
$\{p(r), \rho(r), \nu(r)\}$ satisfying one equation, that is, the conservation equation (\ref{con1}) . Hence to obtain a
solution we must add additional information. However it is not
clear what kind of restriction should be considered to close the
system in the brane. As mentioned early, it is well
known that the non closure of the braneworld equations is directly
related with the projection of the bulk Weyl tensor ${\cal
E}_{\mu\nu}$ on the brane. This is an open problem for which the
solution requires  more information of the bulk geometry and a
better understanding of how our 4D spacetime is embedded in the
bulk.

To clarify the way to obtain some criterion which helps in
searching a solution of the problem described in the previous paragraph, let us start with the
apparent simplest way to find a solution to the system of equations in the brane: using (\ref{uu}) in the original form of the equation (\ref{usual}), which is the field equation
\begin{equation}
\label{ec1}-8\pi \left( \rho
+\frac{1}{\sigma}\left(\frac{\rho^2}{2}+\frac{6}{k^4}\cal{U}\right)
\right) =-\frac 1{r^2}+e^{-\lambda }\left( \frac
1{r^2}-\frac{\lambda _1}r\right),
\end{equation}
we have a
first order linear differential equation to the geometric function $e^{-\lambda }$, given by
\begin{eqnarray}
\label{e1g}
-\lambda_1e^{-\lambda}+e^{-\lambda}\left(\frac{\nu_{11}+\nu_1^2/2+2\nu_1/r+2/r^2}{\nu_1/2+2/r}\right)=\frac{2}{r^2(\nu_1/2+2/r)}
\nonumber \\
\nonumber \\
-8\pi\frac{\left(\rho-3p-\frac{1}{\sigma}\rho
(\rho+3p)\right)}{(\nu_1/2+2/r)},
\end{eqnarray}
which formal solution is
\begin{eqnarray}
\label{primsol}
e^{-\lambda}=e^{-I}\left(\int_0^r\frac{e^I}{(\frac{\nu_1}{2}+\frac{2}{r})}\left[\frac{2}{r^2}-8\pi(\rho-3p-\frac{1}{\sigma}\left(\rho^2+3\rho
p)\right)\right]dr+c\right),
\end{eqnarray}
with
\begin{eqnarray}
\label{I} I\equiv
\int\frac{(\nu_{11}+\frac{\nu_1^2}{2}+\frac{2\nu_1}{r}+\frac{2}{r^2})}{(\frac{\nu_1}{2}+\frac{2}{r})}dr.
\end{eqnarray}
Then when a solution $\{p$,
$\rho$, $\nu\}$ to (\ref{con1}) is found, we would be able to find
$\lambda$, ${\cal P}$ and ${\cal U}$ by (\ref{primsol}),
(\ref{pp}) and (\ref{uu}) respectively. Therefore, from the point
of view of a brane observer, finding a solution in the brane, at least from the mathematical point of
view, seems not very complicated. However it was shown in
\cite{jovalle07} that finding a consistent solution by starting from any arbitrary solution $\{p,\rho,\nu\}$ to the
conservation equation (\ref{con1}), in general does not lead to a
solution for $e^{-\lambda}$ having the expected form, which is
\begin{eqnarray}
\label{expect}  e^{-\lambda}=1-\frac{8\pi}{r}\int_0^r r^2\rho
dr+\frac{1}{\sigma}(Bulk\;\;effects).
\end{eqnarray}
If the solution found to $e^{-\lambda}$ cannot be written by the
way given by (\ref{expect}), then the general relativity limit,
given through $\frac{1}{\sigma}\rightarrow 0$, will not be
regained. Unfortunately this happen when we start from any
arbitrary solution $\{p,\rho,\nu\}$ to (\ref{con1}).  The source of
this problem has to do with the formal solution to $e^{-\lambda}$
given through (\ref{primsol}). Such a solution has mixed general
relativity terms with non-local bulk terms in such a way that
makes impossible to regain general relativity.


A different way to explain why the formal solution (\ref{primsol})
leads to the so-called "general relativity limit problem" is
detailed next. First of all, it seems make sense to consider a
solution to the geometric function $\lambda$ as a generalization of the standard
general relativity solution through
\begin{equation}
\label{usual2} e^{-\lambda}=1-\frac{8\pi}{r}\int_0^r
r^2\tilde{\rho}dr,
\end{equation}
where
\begin{equation}
\label{deneffec} \tilde{\rho}\equiv\rho
+\frac{1}{\sigma}\left(\frac{\rho^2}{2}+\frac{6}{k^4}\cal{U}\right)
\end{equation}
is the effective density having local and non-local bulk effects
on the brane. It can be seen that taking the limit
$\frac{1}{\sigma}\rightarrow 0$ the well known general relativity
solution is regained. Thus the lost of the general relativity
seems not being a problem anymore. However we will see that the
naive solution (\ref{usual2}) is not a true solution at all. Let us
start using (\ref{uu}) in (\ref{deneffec}) to obtain
\begin{equation}
\label{deneffec2} \tilde{\rho}=\rho
-\frac{1}{\sigma}\left(\rho^2+3\rho
p\right)+\frac{1}{8\pi}\left(2G_2^2+G_1^1\right)-3p,
\end{equation}
thus (\ref{usual2}) is written as
\begin{equation}
\label{usual3} e^{-\lambda}=1-\frac{8\pi}{r}\int_0^r
r^2{\rho}dr+\frac{8\pi}{r}\int_0^r
r^2\left[-\frac{1}{\sigma}\left(\rho^2+3\rho
p\right)+\frac{1}{8\pi}\left(2G_2^2+G_1^1\right)-3p\right]dr.
\end{equation}
Thus we can see that the "solution" (\ref{usual2}) depends itself
on $\lambda(r)$ and $\lambda_1(r)$ through $G_1^1$ and $G_2^2$,
hence it represents an integral differential equation for the
geometrical function $\lambda(r)$, something completely different
from the general relativity case, and a direct consequence of the
non-locality of the braneworld equations. A way to get it over
will be explained in the next section.

\section{Generating a constraint in the brane}

So far we have two problems closed related making difficult the
study of non-uniform braneworld stars, which any braneworld
observer has to face, namely, the non possibility of regaining
general relativity when the formal solution (\ref{primsol}) is
implemented or the existence of an integral differential equation
when the standard solution (\ref{usual2}) is enforced. A method
already presented in \cite{jovalle07} to overcome these two
problems, which eventually leads to a constraint producing a
reduction of the degrees of freedom in the brane, is explained in
detail next.

Fist of all let us split the differential equation (\ref{e1g}) as
\begin{eqnarray} \label{edlrw}
\left[\frac{-\lambda_1e^{-\lambda}}{r}+\frac{e^{-\lambda}}{r^2}-\frac{1}{r^2}+8\pi
\rho\right]+\left[-\lambda_1e^{-\lambda}(\frac{\nu_1}{2}+\frac{1}{r})+\right.
\nonumber \\
\nonumber \\
\left.e^{-\lambda}(\nu_{11}+\frac{\nu_1^2}{2}+2\frac{\nu_1}{r}+\frac{1}{r^2})-\frac{1}{r^2}-8\pi
3p-\frac{8\pi}{\sigma}\rho\left(\rho+3p\right)\right]=0,
\end{eqnarray}
here the left bracket has the standard general relativistic terms
and the right bracket the bulk effects which modify the general
relativistic equation. It can be seen that not all terms in the
right bracket are manifestly bulk contributions, that is, not all
of them are proportional to $\frac{1}{\sigma}$. Indeed only high
energy terms are manifestly bulk contributions, while terms whose
source is Weyl curvature remain as not explicit bulk contribution.
This non-local terms, which are non explicit bulk contributions,
are easily mixed with general relativistic terms, as is shown in
the differential equation (\ref{e1g}). Hence the solution
eventually found to $e^{-\lambda}$, when (\ref{e1g}) is solved, is
written in such a way that will never have the form shown in
(\ref{expect}). Therefore it will not be possible to regain the
general relativistic solution when the $\frac{1}{\sigma}\rightarrow
0$ limit is taken.

Keeping in mind general relativity as a limit, and the fact that a braneworld observer should see a geometric deformation due to five dimensional gravity effects, the following solution is proposed for the radial metric component:
\begin{eqnarray}
\label{expectg0}  e^{-\lambda}=\mu+Geometric\;\;Deformation,
\end{eqnarray}
where the $\mu$ function is the well known general relativistic
solution, given by
\begin{eqnarray}
\label{grsolution}  \mu =1-\frac{8\pi}{r}\int_0^r r^2\rho dr.
\end{eqnarray}
The unknown {\it geometric deformation} in (\ref{expectg0}) should have two sources: extrinsic curvature and five dimensional Weyl curvature, hence it can be written as a generic $f$ function
\begin{eqnarray}
\label{expectg}  e^{-\lambda}=\mu+f
\end{eqnarray}
which at the end will have the form
\begin{equation}
\label{deform}
f= \frac{1}{\sigma}(\;high\;energy\;terms)\; + \;non\;local\;terms,
\end{equation}
where, according to (\ref{usual3}), the non local terms in (\ref{deform}) must be related with the anisotropy projected onto the brane.

Demanding that the proposed solution (\ref{expectg}) satisfies
(\ref{edlrw}), a first order differential equation to $f$ is
obtained, given by
\begin{equation}
\label{diffeqtof}
f_1+\left[\frac{\nu_{11}+\frac{\nu_1^2}{2}+\frac{2\nu_1}{r}+\frac{2}{r^2}}{\frac{\nu_1}{2}+\frac{2}{r}}\right]f=\frac{
\frac{8\pi}{\sigma}\rho(\rho+3p)-H(p,\rho,\nu)}{\frac{\nu_1}{2}+\frac{2}{r}},
\end{equation}
where the function $H(p,\rho,\nu)$ is defined as
\begin{equation}
\label{H} H(p,\rho,\nu)\equiv
\left[\mu_1(\frac{\nu_1}{2}+\frac{1}{r})+\mu(\nu_{11}+\frac{\nu_1^2}{2}+\frac{2\nu_1}{r}+\frac{1}{r^2})-\frac{1}{r^2}\right]-8\pi
3p.
\end{equation}
Solving (\ref{diffeqtof}) the $f$ function is written as\footnote{
The constant of integration is put equal to zero to avoid a
singular solution.}
\begin{equation}
\label{fsolution}
f=e^{-I}\int_0^r\frac{e^I}{(\frac{\nu_1}{2}+\frac{2}{r})}\left[H(p,\rho,\nu)+\frac{8\pi
}{\sigma}\left(\rho^2+3\rho p\right)\right]dr,
\end{equation}
where local and non-local bulk effects can be seen. In (\ref{fsolution}) the function $I$ is given again by the expression shown in (\ref{I}).

It is easy to see through the equations
(\ref{usual})-(\ref{uu}) evaluated at $\sigma^{-1}=0$ (general
relativity) that the non-local function $H(p,\rho,\nu)$ can be
written as
\begin{equation}
\label{H2}
H(p,\rho,\nu)=\left(2G_2^2+G_1^1\right)|_{\sigma^{-1}=0}-8\pi 3p,
\end{equation}
which clearly correspond to an anisotropic term. In the general
relativity case, that is, perfect fluid solution, the function
$H(p,\rho,\nu)$ vanishes as a consequence of the isotropy of the
solution. However, in the braneworld case, the general relativity
isotropic condition $H(p,\rho,\nu)=0$, in general should not be
satisfied anymore. There is not reason to believe that the
modifications undergone by $p$, $\rho$ and $\nu$, due to the bulk
effects on the brane, do not modify the isotropic condition
$H(p,\rho,\nu)=0$. Therefore in general we have $H(p,\rho,\nu)\neq
0$. Thus the solution to the geometric function $\lambda(r)$ is
finally written by
\begin{equation}
\label{edlrwss} e^{-\lambda}={1-\frac{8\pi}{r}\int_0^r r^2\rho
dr}+e^{-I}\int_0^r\frac{e^I}{(\frac{\nu_1}{2}+\frac{2}{r})}\left[H(p,\rho,\nu)+\frac{8\pi
}{\sigma}\left(\rho^2+3\rho p\right)\right]dr.
\end{equation}

In order to recover general relativity, the following condition
must be satisfied
\begin{equation}
\label{constraintf} lim_{{\sigma}^{-1}\rightarrow\;\;
0}\;\;\int_0^r\frac{e^I}{(\frac{\nu_1}{2}+\frac{2}{r})}H(p,\rho,\nu)dr
=0.
\end{equation}
The expression (\ref{constraintf}) can be interpreted as a
constraint whose physical meaning is nothing but the necessary
condition to regain general relativity. A simple solution to
(\ref{constraintf}) which has a direct physical interpretation is
\begin{equation}
\label{constraint2} H(p,\rho,\nu)=0.
\end{equation}
The constraint (\ref{constraint2}) explicitly ensure the general
relativity limit\footnote{Indeed, every perfect fluid solution
satisfies H=0 (see next section) } through the solution
(\ref{edlrwss}), and has been proven to be useful in finding
solutions which posses general relativity as a limit
\cite{jovalle07}. It was clearly established through the equation
(\ref{H2}) that the constraint (\ref{constraint2}) represents a
condition of isotropy in general relativity. Thus in the context
of the braneworld the constraint (\ref{constraint2}) has a direct
physical interpretation: {\it eventual bulk corrections to $p$,
$\rho$ and $\nu$ will not produce anisotropic effects on the
brane.} A clearer physical meaning of the constraint (\ref{constraint2}) will be illustrated in the next section.

\section{Generating the braneworld version of any known general relativistic solution}

After the constraint (\ref{constraint2}) is imposed, the problem
on the brane is reduced to finding a solution
$\{p(r),\rho(r),\nu(r)\}$ satisfying (\ref{con1}) and
(\ref{constraint2}). The author found that the following general
expression for the geometric variable $\nu(r)$
\begin{equation}
\label{nuused} e^\nu=A(1+Cr^m)^n
\end{equation}
produces an analytic expression for the integral shown in (\ref{I}) which has to be used in
(\ref{edlrwss}), and a complicated integral equation for $\rho$
when (\ref{nuused}) is used in (\ref{con1}) and
(\ref{constraint2}). It is difficult to figure out appropriates
values for the set of constants $\{A, C, m, n\}$ leading to exact
expressions for $p$ and $\rho$, and even more in the searching of
exact and physically acceptable solutions. Nevertheless exact
solutions for $\{p(r),\rho(r),\nu(r)\}$ where found in Ref.
\cite{jovalle07} and an exact solution for the complete system
$\{p(r),\rho(r),\nu(r),\lambda(r),{\cal U}(r), {\cal P}(r)\}$ is
reported in Ref. \cite{jovalle207}. It is worth noticing that in
the case of a uniform distribution the system is {\it closed},
thus it is not necessary to impose any additional restriction
except the constraint (\ref{constraint2}), which will produce a
non lineal differential equation for the geometric function $\nu$.

The way described in the previous paragraph was successfully used
by the author. However now we become aware of an important issue
regarding the constraint (\ref{constraint2}), which was not
previously mentioned: {\it every solution for a perfect fluid
satisfies the minimal anisotropic condition represented through
the constraint H=0}. Therefore, the constraint (\ref{constraint2})
represents a natural way to generalize perfect fluid solutions
(general relativity) in the context of the braneworld. The method
consist in taking a known perfect fluid solution
$\{p(r),\rho(r),\nu(r)\}$, then $\lambda(r)$, ${\cal P}(r)$ and
${\cal U}(r)$ are found through (\ref{edlrwss}), (\ref{pp}) and
(\ref{uu}) respectively. Of course the fact that the constraint
(\ref{constraint2}) be satisfied by every known perfect fluid
solution does not mean that any of these solutions can be directly
used to find an exact braneworld solution. In order to investigate
if one particular general relativistic solution has an exact
braneworld version, the first step is to analyze the temporal
component of the metric. If $\nu(r)$ is not simple enough, then
hardly it will provide an analytic expression for (\ref{I}). Even
in the case where (\ref{I}) be analytic, finding an exact
expression for $\lambda(r)$ using (\ref{edlrwss}) is very
difficult. Nevertheless, this approach, which is entirely based in
the constraint $H=0$, always can be used to obtain the braneworld
version of \textit{any} general relativistic solution by numerical
methods. Next it is explained a direct physical meaning of this constraint.

Using the geometric deformation $f$ as shown in Eq. (\ref{expectg}) in the expression for ${\cal P}$ given in (\ref{pp}), it is found that the anisotropy induced by five dimensional effects may be written in term of the geometric deformation as
\begin{eqnarray}
\label{ppf}
\frac{48\pi}{k^4}{\cal P}&=&-\frac{1}{r^2}+\mu(\frac{1}{r^2}+\frac{\nu_1}{r})-\frac{1}{4}\mu(2\nu_{11}+\nu_{1}^2+2\frac{\nu_1}{r})-\frac{1}{4}\mu_1(\nu_1+\frac{2}{r})
\nonumber \\
&&+f(\frac{1}{r^2}+\frac{\nu_1}{r})-\frac{1}{4}f(2\nu_{11}+\nu_{1}^2+2\frac{\nu_1}{r})-\frac{1}{4}f_1(\nu_1+\frac{2}{r}).
\end{eqnarray}

When the constraint (\ref{constraint2}) is imposed the anisotropy induced shown in (\ref{ppf}) is written in terms of the geometric deformation by
\begin{eqnarray}
\label{ppf2}
\frac{48\pi}{k^4}{\cal P}&=&
\frac{3}{2}\left[-\frac{1}{r^2}+\mu(\frac{1}{r^2}+\frac{\nu_1}{r})-8\pi{p}\right]
\nonumber\\
&&+f^{*}(\frac{1}{r^2}+\frac{\nu_1}{r})-\frac{1}{4}f^{*}(2\nu_{11}+\nu_{1}^2+2\frac{\nu_1}{r})-\frac{1}{4}f^{*}_1(\nu_1+\frac{2}{r}),
\end{eqnarray}
where
\begin{equation}
\label{mindef}
f^{*}=\frac{1}{\sigma}(\;high\;energy\;terms)\; + \underbrace{\;non\;local\;terms}_{=\,0}
\end{equation}
is the {\it minimal geometric deformation}, whose explicit form may be seen by (\ref{fsolution}) as following
\begin{equation}
\label{fsolutionmin}
f^{*}=\frac{8\pi
}{\sigma}e^{-I}\int_0^r\frac{e^I}{(\frac{\nu_1}{2}+\frac{2}{r})}\left(\rho^2+3\rho p\right)dr.
\end{equation}
The expression (\ref{fsolutionmin}) represents a minimal geometric deformation in the sense that all sources of the geometric deformation $f$ have been removed except those produced by the density and pressure, which are always present in a stellar distribution\footnote{An even minimal deformation is obtained for a dust cloud, where $p=0.$}. It is clear that the geometric deformation represented by the $f^{*}$ function is a source of anisotropy, as may be seen in (\ref{ppf2}). However there is another source for ${\cal P}$, which is represented by the bracket shown in (\ref{ppf2}). Nevertheless, when the constraint (\ref{constraint2}) is imposed the bracket shown in (\ref{ppf2}) should be zero, since eventual bulk corrections to $p$, $\rho$ and $\nu$ do not produce anisotropic effects on the brane. Indeed, every general relativistic solution produces
\begin{eqnarray}
\label{ppf3}
\frac{48\pi}{k^4}{\cal P}&=&
\frac{3}{2} \underbrace{\left[-\frac{1}{r^2}+\mu(\frac{1}{r^2}+\frac{\nu_1}{r})-8\pi{p}\right]}_{-8\pi{T}_{1}^{1}-G_{1}^{1}=\,0}
\nonumber\\
&&+f^{*}(\frac{1}{r^2}+\frac{\nu_1}{r})-\frac{1}{4}f^{*}(2\nu_{11}+\nu_{1}^2+2\frac{\nu_1}{r})-\frac{1}{4}f^{*}_1(\nu_1+\frac{2}{r}),
\end{eqnarray}
thus leaving the minimal geometric deformation $f^{*}$ as the only source for the anisotropy induced inside the stellar distribution. Therefore every general relativistic solution satisfies the minimal anisotropic effect onto the brane. In consequence any general relativistic solution may be used to obtain a consistent braneworld stellar solution. Here below is shown the basic steps to use this approach.
\begin{itemize}

\begin{figure}
\includegraphics[width=120mm]{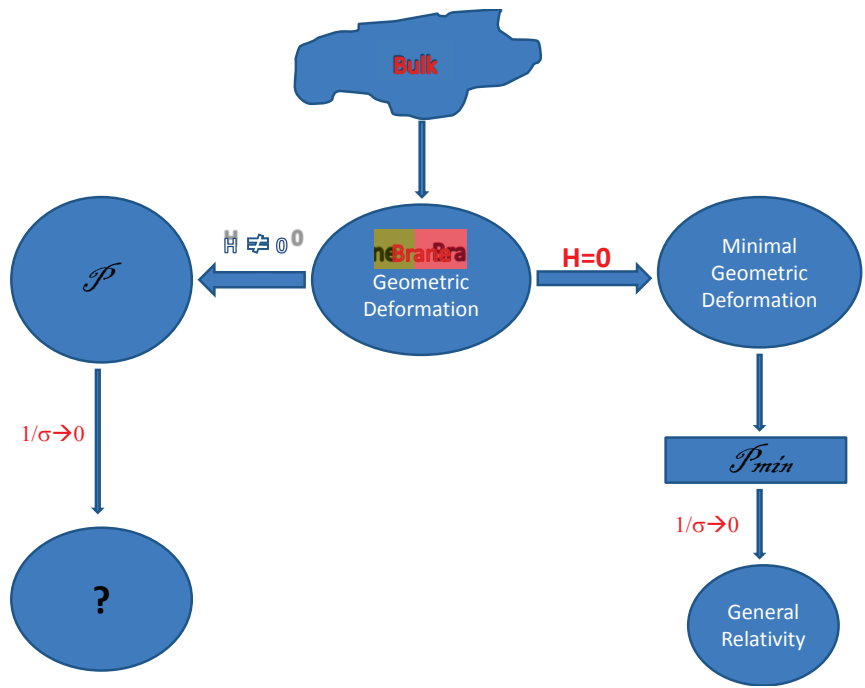}
\caption{The bulk effects produce a geometric deformation onto the brane, which is minimal when $H=0$ is imposed. Thus the anisotropic consequence onto the brane is minimal as well.}
\end{figure}

\item Step 1: Impose the constraint $H(p,\rho,\nu)=0$ to make sure
we have a solution for the geometric function $\lambda(r)$ with the
correct limit:
\begin{eqnarray}
e^{-\lambda(r)}={1-\frac{8\pi}{r}\int_0^r r^2\rho
dr}+{\frac{8\pi
}{\sigma}e^{-I}\int_0^r\frac{e^I}{(\frac{\nu'}{2}+\frac{2}{r})}\left(\rho^2+3\rho
p\right)dr}.\nonumber
\end{eqnarray}

\item Step 2: Pick a known general relativistic solution
$(p,\rho,\nu)$ to the conservation equation
$\;p'=-\frac{\nu'}{2}(\rho+p)\;$.

\item Step 3: Find ${\cal P}$ and ${\cal U}$ by equations shown in (\ref{pp}) and (\ref{uu}).

\item Step 4: Drop out the condition of vanishing pressure at the
surface to obtain the bulk effect on any constant $C\rightarrow
C(\sigma)$. Then we are able to find the bulk effect on pressure and
density.

\end{itemize}

In the next section the Schwarzschild's interior solution will be considered as a clear example of this approach.

\section{Conclusions and outlook}

In the context of the Randall-Sundrum braneworld, a detailed description of the approach based in the
minimal anisotropic consequence onto the brane was carefully presented. The explicit form of the anisotropic stress was obtained in terms of the geometric deformation undergone by the radial metric component, thus showing the role played by this deformation as a source of anisotropy inside the stellar distribution. It was shown that this geometric deformation is minimal when a general relativistic solution is considered, therefore any general relativistic solution belongs to a subset of braneworld solutions producing a minimal anisotropic consequence onto the brane. It was found that through this approach, it is possible to generate the braneworld version of any known general relativistic solution, thus overcoming the non-closure problem of the braneworld equations.

In this work a consistent way to generate the braneworld version of any general relativistic solution in the context of the Randall-Sundrum theory was shown. This approach might be extended in the case of braneworld theories without $Z_2$ symmetry or any junction conditions, as those introduced in \cite{maia2004} and \cite{maia2005}, which have been successfully used in the astrophysics context \cite{sepangi2009}. Another subjects of interest is the use of this approach in brane theories with variable tension, as introduced by \cite{gergely2009} in the cosmological context, and the study of codimension-2 braneworld theories, as those developed in \cite{minas2008} and \cite{papa2008}. A possible extension of this approach in all these theories is currently being investigated.

\section*{Acknowledgments}

This work was supported by {\bf Decanato de Investigaci\'on y Desarrollo, USB}. Grant:
S1-IN-CB-002-09, and by {\bf FONACIT}. Grant: S2-2009000298.




\end{document}